\RequirePackage{ifpdf}
\ifpdf 
\documentclass[pdftex]{sigma}
\else
\documentclass{sigma}
\fi

\begin{document}

\allowdisplaybreaks

\renewcommand{\thefootnote}{$\star$}

\renewcommand{\PaperNumber}{086}

\FirstPageHeading

\ShortArticleName{Time Asymmetric Quantum Mechanics}

\ArticleName{Time Asymmetric Quantum Mechanics\footnote{This
paper is a contribution to the Proceedings of the Workshop ``Supersymmetric Quantum Mechanics and Spectral Design'' (July 18--30, 2010, Benasque, Spain). The full collection
is available at
\href{http://www.emis.de/journals/SIGMA/SUSYQM2010.html}{http://www.emis.de/journals/SIGMA/SUSYQM2010.html}}}

\Author{Arno R.~BOHM~$^\dag$, Manuel GADELLA~$^\ddag$ and Piotr KIELANOWSKI $^\S$}

\AuthorNameForHeading{A.R.~Bohm, M.~Gadella and P.~Kielanowski}

\Address{$^\dag$~Department of Physics, University of Texas at
Austin, Austin, TX 78712, USA}
\EmailD{\href{mailto:bohm@physics.utexas.edu}{bohm@physics.utexas.edu}}

\Address{$^\ddag$~Departamento de FTAO, Universidad de Valladolid,
47071 Valladolid, Spain}
\EmailD{\href{mailto:manuelgadella1@gmail.com}{manuelgadella1@gmail.com}}

\Address{$^\S$~Cinvestav, Dept Fis, Mexico City 07000, DF Mexico}
\EmailD{\href{mailto:kiel@physics.utexas.edu}{kiel@physics.utexas.edu}}

\ArticleDates{Received January 30, 2011, in f\/inal form August 22, 2011;  Published online September 03, 2011}

\Abstract{The meaning of time asymmetry in quantum physics is
  discussed. On the basis of a mathematical theorem, the
  Stone--von~Neumann theorem, the solutions of the dyna\-mi\-cal equations,
  the Schr\"odinger equation~\eqref{4} for states or the Heisenberg
  equation~\eqref{eq:15a} for observables are given by a unitary group. Dirac
  kets require the concept of a RHS (rigged Hilbert space) of Schwartz
  functions; for this kind of RHS a mathematical theorem also leads to
  time symmetric group evolution. Scattering theory suggests to
  distinguish mathematically between states (def\/ined by a preparation
  apparatus) and observables (def\/ined by a registration apparatus
  (detector)). If one requires that scattering resonances of width
  $\Gamma$ and exponentially decaying states of lifetime
  $\tau=\frac{\hbar}{\Gamma}$ should be the same physical entities
  (for which there is suf\/f\/icient evidence) one is led to a pair of
  RHS's of Hardy functions and connected with it, to a semigroup time
  evolution $t_{0}\leq t<\infty$, with the puzzling result that there
  is a quantum mechanical beginning of time, just like the big bang
  time for the universe, when it was a quantum system. The decay of
  quasi-stable particles is used to illustrate this quantum mechanical
  time asymmetry. From the analysis of these processes, we show that
  the properties of rigged Hilbert spaces of Hardy functions are
  suitable for a formulation of time asymmetry in quantum mechanics.}


\Classification{81Q65}

\renewcommand{\thefootnote}{\arabic{footnote}}
\setcounter{footnote}{0}

\section{Introduction}

In this article\footnote{This manuscript is based on lectures given
by A.R.~Bohm at the Centro de Ciencias de Benasque, Spain.}, we
describe the notion of time asymmetry (TA) in quantum mechanics,
give a brief account of its foundations and formulation and present
its relation to the resonance phenomenon.

Time symmetry in quantum mechanics means that the time evolution of
states and ob\-ser\-vables is governed by the unitary group. If we
assume that the Hamiltonian $H$ does not depend explicitly on time,
then this unitary group is provided by the Stone theorem as
$U(t)=e^{-itH/\hbar}$ with $-\infty<t<\infty$. This situation can be
derived from the Hilbert space axiom, which in particular leads to
a unitary group of bounded operators: as a consequence of
this Hilbert space ``boundary condition'' for the Schr\"odinger
equation one obtains the unitary group solution for the state.

If in $U(t)$ the values of $t$ extend from $0\le t<\infty$, the
evolution is time asymmetric. In this case, time evolution for pure
states is not governed by a group. We shall show that in place of
the unitary group one has two semigroups. One semigroup with $t\le
0$ is called the past semigroup, the other, with $t\ge 0$, is called
the future semigroup. The physical nature of these semigroups will
be discussed in the present paper.

Time evolution governed by a semigroup is not possible in a
formulation of quantum mechanics in Hilbert space. The Stone theorem
for the solution of the Schr\"odinger equation and similar for the
solution of the Heisenberg equation automatically leads to groups of
time evolution in Hilbert space, if the Hamiltonian is self adjoint.
Therefore, we have to drop this ``Hilbert space boundary condition''
and replace it by another type of boundary condition\footnote {Or
  another type of paradigm.}.

A new boundary condition for the dynamical equations is obtained by
replacing the Hilbert space by other spaces. This will be the pair
of Hardy spaces,  ${\cal H}_+^2$ for the Heisenberg equation of the
observables and  ${\cal H}_-^2$ for the Schr\"odinger equation. This is
a pair of ``time asymmetric boundary conditions''. Taking into
account that we need to extend the Hilbert space into a rigged
Hilbert space to implement this time asymmetry, we may call the new
boundary condition the ``rigged Hilbert space boundary condition''.
We intend to clarify these notions in the present paper.

The solutions of the Schr\"odinger equation under the Hilbert space
boundary condition
\begin{gather}\label{4}
 i \hbar \frac{d \phi(t)}{d t} = H \phi(t)
\end{gather}
are given by the unitary group
\begin{gather*}
\phi(t)=e^{-i H t/\hbar}\phi(0) ,\qquad \text{for all $t$:}  \quad
-\infty < t < +\infty .
\end{gather*}
This follows from the Stone--von Neumann theorem for the
Schr\"odinger equation~(\ref{4}). Thus in absence of an interaction
with an external reservoir no TA is allowed under Hilbert space
boundary conditions, i.e., the ``Hilbert space boundary condition''
does not allow TA.

TA in quantum physics manifests itself in processes like the decay
of a quasi-stationary state. TA in these processes is suggested by
the fact that the creation or formation of a quasi-stationary state
is not the time reversal of its decay. Formation requires the
realization of some initial conditions~\cite{TDL}, while the decay
is usually spontaneous. Thus, decay can be viewed as an independent
process. This process is time asymmetric as we shall discuss later.
Both processes (formation as well as decay) are time reversal
invariant (except for minute part in $K^{0}$ decay). Resonances are
usually produced by a Hamiltonian pair $\{H_0,H=H_0+V\}$ where both
the ``free'' Hamiltonian $H_0$ and the interaction Hamiltonian $H$
are time reversal invariant, i.e., $[A_T,H_0]=0=[A_T,H]$, where
$A_T$ is the time reversal operator. This means that TA is not to be
mistaken for time reversal non-invariance.

The quasi-stationary state undergoes a decay process for which time
evolution is def\/ined for $t\ge 0$ only. The quasi-stationary state
can be described by a state vector, the Gamow vector which is not
normalizable, it does not admit a f\/inite norm.  Time evolution for a
quasi-stationary state is given by a semigroup with parameter $t\ge
0$. Due to the Stone--von~Neumann theorem this description is not
possible under ``Hilbert space boundary conditions''.

In fact, nature may require boundary conditions other than the
Hilbert space axioms. We suggest that dynamical equations like
equation (\ref{4}) which has time symmetric solutions in Hilbert
space, will also admit time asymmetric solutions under dif\/ferent
time asymmetric boundary conditions. Time asymmetry in quantum
mechanics means time asymmetric boundary conditions for time
symmetric dynamical equation (\ref{4}).

A well known example of a time asymmetric boundary condition for
time symmetric dynamical equations is the radiation arrow of time.
Maxwell's equations (dynamical dif\/ferential equations) are symmetric
in time.  A boundary condition excludes the strictly incoming f\/ields
and selects only the retarded f\/ields of the other sources in the
region:
\begin{gather*}
A^\mu(x)=A^\mu_{\text{ret}}(x)+A^\mu_{\text{in}}(x)=A^\mu_{\text{ret}}(x).
\end{gather*}
The condition $A^\mu_{\text{in}}(x)=0$ is the Sommerfeld radiation condition.
The boundary condition chooses of the two solutions for the Maxwell
equations
\begin{gather*}
A^\mu_{\mp}(\vec{x},t)=
\int \delta\left(t^\prime-\left(t\mp\frac{\vert\vec{x}-\vec{x}\,{}^\prime\vert}{c}\right)\right)
\frac{j^\mu(\vec{x}\,{}^\prime,t^\prime)}{\vert \vec{x}-\vec{x}\,{}^\prime
\vert}\, d^3x^\prime d t^\prime ,
\end{gather*}
only the retarded solution
\begin{gather*}
A^\mu_{\text{ret}}(\vec{x},t)\equiv A^\mu_-(\vec{x},t)=\int \frac{j^\mu
\left(\vec{x}\,{}^\prime,t-\frac{\vert\vec{x}-\vec{x}\,{}^\prime\vert}{c}\right)}
{\vert \vec{x}-\vec{x}\,{}^\prime \vert} d^3 x^\prime .
\end{gather*}

However, in standard quantum mechanics one mostly does not use
time-asymmetric boundary conditions, but instead uses the Hilbert
space axiom for the solutions of the Schr\"odinger (or Heisenberg)
equation. To obtain time asymmetric boundary conditions in quantum
theory, we need to revise some basic ideas about the boundary
conditions for states. The same should be the case for observables
in the Heisenberg representation. These ideas
\cite{B_I,B1,B2,B3,B4,B5} are the subject of the next section.

\section{The notions of state and observable revisited}\label{section2}

Let us consider a scattering experiment in the non-relativistic
case. In a scattering experiment, a state is ``created'' or prepared
by a {\it preparation apparatus}. This state interacts with a center
of forces and as the result of this interaction a new state emerges
which can be observed in the distant future  by a {\it registration
apparatus}. In this context \cite{B_I}, we have to def\/ine the
notions of state and observable in the following way:
\begin{center}
  \begin{minipage}{0.9\linewidth}
    \emph{States} are described by  a density operator $\rho$ or by a
state vector $\phi$.

\emph{Observables} are described by selfadjoint operators
$A=A^{\dagger}$, e.g., by a projection ope\-ra\-tor
$\Lambda=\Lambda^{2}$; if an observable is given by
$\Lambda=\vert\psi\rangle\langle\psi\vert$ then $\psi$ (and every
$e^{i \alpha}\psi$, $\alpha$ real) will be called an
\textit{observable vector} with the property $\Lambda$.
\end{minipage}
\end{center}
States {\it are prepared by} a preparation apparatus, like an
accelerator. Thus, in a scattering expe\-ri\-ment a state is identif\/ied
with what is usually called an {\it incoming state}. On the other
hand, an observable {\it is registered by} a registration apparatus,
e.g., a detector. Detected {\it outgoing states} are really
observables $|\psi(t)\rangle\langle\psi(t)|$ and observables obey
the Heisenberg equation of motion. The reason for this is that the
experimental quantities are the Born probabilities to measure an
observable $\Lambda$ in the state $\rho$ described by ${\rm
Tr}\,(\Lambda\rho)$. They are calculated in theory as the Born
probabilities and they are measured as the ratio of large numbers of
detector counts $N(t)/N$. In the Schr\"odinger picture they are
given by:
\begin{gather*}
 \frac{N(t)}{N}\approx \mathcal{P}_{\rho(t)}(\Lambda)\equiv
 {\rm Tr }\,(\Lambda_{0} \rho(t))= |\langle
 \psi^-|\phi^+(t)\rangle|^2 .
\end{gather*}
The signs $+$ and $-$ are chosen in agreement with signs as they
will appear in the Lippmann--Schwinger equations in Section~\ref{section3.1}. In
particular, this means that $\phi^+=\Omega_{\rm IN}\,\phi^{\rm in}$
and $\psi^-=\Omega_{\rm OUT}\,\psi^{\rm out}$, where ``in'' and
``out'' means incoming and outgoing respectively.

In the Heisenberg picture the Born probabilities are given by:
\begin{gather*}
\frac{N(t)}{N}\approx \mathcal{P}_{\rho}(\Lambda(t))\equiv {\rm Tr
}\,(\Lambda(t) \rho_0)= |\langle \psi^-(t)|\phi^+\rangle|^2 .
\end{gather*}
with time dependent observable $\Lambda(t)$ or
$\vert\psi(t)\rangle\langle\psi(t)\vert$. The sign $\approx$
indicates that these equali\-ties are the comparison between the
experimental quantity, the counting rates $N(t)/N$, and the
calculated, theoretical Born probabilities. The equivalence between
the Schr\"odinger and Heisenberg pictures is given by the following
mathematical identity:
\begin{gather*}
\operatorname{Tr}\,(|\psi^-\rangle\langle
\psi^-|\phi^+(t)\rangle\langle\phi^+(t)|)=|\langle
\psi^-|\phi^+(t)\rangle|^2= |\langle \psi^-(t)|\phi^+\rangle|^2 =
{\rm Tr }\, (
|\psi(t)\rangle\langle\psi(t)|\phi\rangle\langle\phi|  )  .
\end{gather*}

We ask the following question: is there any evidence that the time
$t$ may take any value? This means, is there any evidence that
$U(t)=e^{iHt/\hbar}$ in the Heisenberg picture and $e^{-iHt/\hbar}$
in the Schr\"odinger picture makes sense for any value of $t$ in
$(-\infty,\infty)$? It seems obvious that, in analogy to the
radiation arrow of time, a state $\phi$ must be prepared at a time
$t_0$ before the observable $|\psi(t)\rangle\langle\psi(t)|$ can be
measured in it. This property is a form of \textbf{causality
principle}, e.g., {\it the
  detector cannot count the decay products
  $\vert\psi(t)\rangle\langle\psi(t)\vert$ before the decaying state
  $\vert\phi\rangle\langle\phi\vert$ has been prepared}.

In this sense,  we have a \textbf{Quantum Mechanical Arrow of Time}.
The Born probability to measure the observable
$|\psi(t)\rangle\langle\psi(t)|$ in the state $\phi$,
\begin{gather*}
  \mathcal{P}_\phi(\psi(t))=|\langle\psi(t)|\phi\rangle|^2=|\langle
  e^{iHt/\hbar}\psi|\phi\rangle|^2=|\langle\psi|e^{-iHt/\hbar}
  \phi\rangle|^2=|\langle\psi|\phi(t)
  \rangle|^2={\cal P}_{\phi(t)}(\psi),
\end{gather*}
exists (experimentally) only for $t\geq t_0$ $(=0)$ where $t_0$ is
the time at which the state has been prepared and the observable can
be detected or ``registered'', i.e., $t_0$ is the preparation time of
the state $\phi$.

In contrast, the Hilbert space axiom (of conventional QM) predicts
$|\langle\psi(t)|\phi\rangle|^2$ for all\linebreak \mbox{$-\infty < t <
  +\infty$}, by the Stone theorem.

In order to obey \textbf{the causality principle} we have to f\/ind a
theory for which the solutions of the Schr\"odinger equation
$\phi(t)$ evolve by a \textit{semigroup} $\mathcal{U}_-(t)=e^{-iH_-
t/\hbar}$, $t_0=0\leq t < \infty$, or for which the solutions of the
Heisenberg equation, $\psi(t)$, evolve by a \textit{semigroup}
$\mathcal{U}_+(t)=e^{+i H_+
  t/\hbar}$, $0\leq t < \infty$. By $H_-$ and $H_+$ we denote the
generators of the semigroups ${\mathcal U}_-(t)$, $0\le t<\infty$, and~${\mathcal U}_+(t)$, $0\le t<\infty$, respectively.

From a mathematical point of view ${\mathcal U}_\pm^\times$ are the
extensions of the Schr\"odinger evolution operator beyond Hilbert
space to the duals of two rigged Hilbert spaces. This extensions exist
only for $0\leq t < \infty$. Analogously, $H_\pm^{\times}$ are the
respective extensions of the Hilbert space Hamiltonian~$H$. These
extensions will be introduced in Section~\ref{section3.3}.

Then, the Born probability for the observable
$\vert\psi\rangle\langle\psi\vert$ in the state $\phi$ is predicted
in this time asymmetric theory for $t\ge t_0=0$ only
\begin{gather}\label{18}
  \mathcal{P}_\phi(\psi(t))=|\langle\psi|\phi(t)\rangle|^2
  =|\langle\psi|e^{-iHt/\hbar}\phi\rangle|^2=|\langle
  e^{iHt/\hbar}\psi|\phi\rangle|^2=|\langle\psi(t)|\phi\rangle|^2,
 \!\! \qquad t\ge t_0=0 .\!\!\!
\end{gather}
Here $t_{0}=0$ represents the preparation time of the state $\phi$.
The question is then: can one observe this time $t_0$ and how can
one observe it? As is common for quantum physical values one expects
that $t_0$ is measured by an ensemble of values $\{t_0^{(n)}\}$~\cite{BB}. Two examples are discussed in~\cite{B_I}. These examples
may show evidence that $i)$~states can be prepared during a f\/inite
time; $ii)$~events can be detected individually.

In summary,  the detector cannot count the decay products of a state
before this state has been prepared. This conclusion is the
principle of causality. It means that we have a {\it Quantum
Mechanical Arrow of Time} (QMAT) which  can be formulated as
follows:

{\it The Born probability to measure the observable
  $|\psi(t)\rangle\langle\psi(t)|$ in the state $\phi$ given by ${\cal
    P}_\phi(\psi(t))$ in \eqref{18} exists experimentally only for
  $t\geq t_0$ $(=0)$, where $t_0$ is the preparation time of the
  state~$\phi$.}

The task is to f\/ind a theory for which the solutions of the
Schr\"odinger equation, $\phi(t)$, evolve by a \textit{semigroup}
$\mathcal{U}_-(t)=e^{-iH_- t/\hbar}$, $0\leq t < \infty$ or for
which the solutions of the Heisenberg equation, $\psi(t)$, evolve by
the \textit{semigroup} $\mathcal{U}_+(t)=e^{+i H_+ t/\hbar}$, $0\leq
t < \infty$. Here $H_\pm$  are the generators of the semigroups
$\mathcal{U}_\pm(t)$, respectively. This task will be discussed in
the next section.

\section{From the mathematics for Dirac kets\\ to separate
  representation for states and for observables}\label{section3}

In the previous section we have established the need for time
asymmetry in  quantum mechanics. A preliminary step towards this
formulation of Time Asymmetric Quantum Mechanics (TAQM) is the {\it
Mathematical Theory of Dirac kets using Gel'fand Triplets} based on
the Schwartz space Gel'fand triplet:
\begin{gather}\label{19}
\Phi\subset{\cal H}\subset\Phi^\times .
\end{gather}
Here $\Phi$ is a dense subspace of $\cal H$ endowed with a f\/iner
topology than the topology inherited from $\cal H$, and
$\Phi^\times$ is the (topological) dual of $\Phi$ formed by the
space of continuous {\it antilinear}\footnote{The action
$F\in\Phi^\times$ on a certain $\phi\in\Phi$ is a complex number
denoted by $\langle \phi|F\rangle$. If $f\in\cal H$, it def\/ines an
element of $\Phi^\times$, $F_f$, def\/ined by $\langle
\phi|F_f\rangle=\langle \phi|f\rangle$, where $\langle
\phi|f\rangle$ is the ordinary scalar product on $\cal H$. This
notation is convenient not only because the action of $F$ on $\phi$
extends the scalar product, but also because the scalar product is
antilinear to the left.} functionals on~$\Phi$.

The most popular example of a RHS is given by the Schwartz space
triplet with $\Phi\doteqdot{\cal S}$, then, $\cal H$ is $L^2({\mathbb
  R})$, the space of complex valued Lebesgue square integrable
functions. The dual space of~$\cal S$ is the space of the (antilinear)
tempered distributions~${\Phi}^{\times}\doteqdot\cal S^\times$. This
gives the Dirac ket the mathematical meaning of an antilinear
continuous functional in the Schwartz space, $\vert
E\rangle\in\Phi^{\times}$.

The use of Gel'fand triples \cite{G,M} to implement the Dirac
formulation~\cite{D} of quantum mechanics has been suggested copiously
\cite{B1,B,BG,GG}. Dirac kets are well def\/ined as functionals in
$\Phi^\times$. For certain simple systems (like the free particle or
the harmonic oscillator), the Dirac formulation can be implemented by
using the Schwartz space. The use of the Schwartz space can be applied
to the energy representation~\cite{BG,CG}, in which case the wave
functions are functions of the energy. In this case, the Schwartz
space is used to implement both the space of states~$\{\phi\}$ and the
space of observables $\{\psi\}$, so that $\{\phi\}\equiv
\{\psi\}\equiv\Phi$. In applications to quantum mechanics the values
of energy are bounded from below; the Schwartz space that is used as a
realization of~$\Phi$ is then no longer $\cal S$ but instead the space
of Schwartz functions that vanish on the negative semiaxis:
$\Phi\doteqdot{\cal S}_{{\mathbb R}^+}$. For any pair of functions
$\psi,\phi\in\Phi$ with $\psi(E),\phi(E)\in {\cal S}_{{\mathbb R}^+}$,
we have
\begin{gather}\label{20}
(\psi,\phi)=\int_0^\infty \langle\psi|E\rangle\langle E|\phi\rangle
\,dE , \qquad \psi^*(E)=\langle\psi|E\rangle ,\qquad \phi(E)=\langle
E|\phi\rangle\in {\cal S}_{{\mathbb R}^+}.
\end{gather}

The kets $|E\rangle$ are generalized eigenvectors of the Hamiltonian
with eigenvalue $E$ in the continuous spectrum of $H$,
$\langle\psi\vert H\,|E\rangle=E\langle\psi|E\rangle$ for all
$\psi\in\Phi$. In most cases, the energy (or $H$) does not form a
complete system of commuting observables (csco), so that some
additional observables should be added to $H$ in order to obtain a
csco. These are often the angular momentum and the third component
of the angular momentum, spin, etc. This is the reason why we choose
for the generalized eigevectors of $H$ the kets
$|E,j,j_3,\eta\rangle$, which are generalized eigenvectors of~$H$
and of angular momentum operators $J^{2}$ and $J_{3}$:
$H^{\times}|E,j,j_3,\eta\rangle=E  |E,j,j_3,\eta\rangle$, and of
other observables called $\eta$ here. Here $H^{\times}$ is the
extension of $H$ to $\Phi^{\times}$. In these examples any vector in~$\Phi$ admits the following basis vector expansion in terms of the
generalized eigenvectors of~$H$:
\begin{gather}\label{21}
\phi=\sum_{j,j_3,\eta} \int_0^\infty |E,j,j_3,\eta\rangle\langle
E,j,j_3,\eta|\phi\rangle\,dE .
\end{gather}

In terms of the wave functions, (\ref{21}) implies:
\begin{gather*}
\langle E,j,j_3,\eta|H|\phi\rangle=E\langle
E,j,j_3,\eta|\phi\rangle=E\phi_{j,j_3,\eta}(E),
\end{gather*}
which means that, in the energy representation, the Hamiltonian $H$
is  represented by the multiplication operator, as it should. In the
energy representation, the Gel'fand triplet $\Phi\subset{\cal
H}\subset\Phi^\times$ is represented by ${\cal S}_{{\mathbb
R}^+}\subset L^2({\mathbb R}^+)\subset({\cal S}_{{\mathbb
R}^+})^\times$.

In the energy representation, both states $\phi$ and observables
$|\psi\rangle\langle\psi|$ are given by respective wave functions
$\phi(E)$ and $\psi(E)$ that belong to the Schwartz space ${\cal
  S}_{{\mathbb R}^+}$. Due to a mathematical theorem~\cite[p.~82]{BG}  the time evolution in either Schr\"odinger and
Heisenberg picture extends also to the Schwartz space
triplet~\eqref{19} for all values of $t$: $-\infty<t<\infty$.

However, there is no reason that the set of observables
$|\psi\rangle\langle\psi|$, $A$, and the set of states $\phi$,
$\rho$, should be described by the same space; i.e., that
$\{\phi\}=\{\psi\}$, and that these should be given by the Schwartz
space.

As mentioned above, in scattering experiments one distinguishes
between the in-states $\phi^{+}$ and out-states $\psi^{-}$; but one
is not much concerned with the mathematical properties of the spaces
of in-states $\{\phi^{+}\}$ and of out-states $\{\psi^{-}\}$. The
in-states are prepared by a preparation apparatus (e.g., an
accelerator) and therefore obey the Schr\"odinger equation. The
out-entities are not really states because they are the entities
registered by a detector; therefore the $\psi^{-}$ represent
observables, physically def\/ined by the registration apparatus (e.g.,
a detector). Therefore the~$\psi^{-}$ should obey the
\textit{Heisenberg equation of motion}, i.e.,
\begin{subequations}
  \begin{gather}
    \label{eq:15a}
    i\hbar\frac{\partial}{\partial t}\vert\psi^{-}(t)\rangle
    \langle\psi^{-}(t)\vert=-[H,\vert\psi^{-}(t)\rangle
    \langle\psi^{-}(t)\vert],
  \\
    \label{eq:15b}
    i\hbar\frac{\partial}{\partial t}\psi^{-}=-H\psi^{-}.
  \end{gather}
The time evolution would then be given not by the Schr\"odiger
equation~\eqref{4}, but it would be the time evolution of
observables $A$, i.e., it would be given by
  \begin{gather}
    \label{eq:15c}
    A(t)=\text{e}^{iHt/\hbar}A\text{e}^{-iHt/\hbar} \qquad \text{or} \qquad
    \psi(t)=\text{e}^{iHt/\hbar}\psi.
  \end{gather}
\end{subequations}

Thus we would expect that one needs not one RHS~\eqref{19}, but a pair
of RHS's
\begin{gather}
  \label{eq:16pk}
  \{\psi^{-}\}\equiv\Phi_{+}\subset\mathcal{H}\subset\Phi_{+}^{\times}\qquad \text{and} \qquad
\{\phi^{+}\}\equiv\Phi_{-}\subset\mathcal{H}\subset\Phi_{-}^{\times}.
\end{gather}
One RHS is for the observables $\{\psi^{-}\}$ fulf\/illing the
Heisenberg equation of motion~\eqref{eq:15b}, and the other  RHS is for the
states $\{\phi^{+}\}$ fulf\/illing the Schr\"odinger
equation~\eqref{4}. This idea is supported by the phenomenological
theories of scattering and decay\footnote{The notation
$\phi^{+}\in\Phi_{-}$ and
  $\psi^{-}\in\Phi_{+}$ may appear awkward, it has its origin in the
  dif\/ference of notation in the mathematics literature for Hardy
  spaces~\cite{PK2,T,PW,VW} $\Phi\mp$, and in the physics literature for
  in and out states $\phi^{\pm}$ ($\phi^{\pm}\in\Phi_{\mp}$) of
  scattering theory~\cite{B_I,LS,GMG,GW}.}.

\subsection{From the phenomenological theories of scattering and decay\\
  to a pair of Hardy spaces}\label{section3.1}

To conjecture mathematical properties of the pair of
RHS's~\eqref{eq:16pk} we use as point of departure the in- and out-
plane wave kets $\vert E^{+}\rangle$ and $\vert E^{-}\rangle$ which
fulf\/ill the Lippmann--Schwinger equation\footnote{Here
  $\Omega^-=\Omega_{\rm OUT}$ and $\Omega^+=\Omega_{\rm IN}$.}
\cite{LS,GMG,GW,GG1}:
\begin{gather}\label{23}
\vert E^{\pm}\rangle=\vert E\rangle +
  \frac{1}{E-H\pm i\epsilon}V \vert E\rangle =\Omega^{\pm} \vert
  E\rangle,\qquad \epsilon\to +0.
\end{gather}
The vectors $|E^+\rangle$, $E\in{\mathbb R}^+=[0,\infty)$, are taken
as a basis systems for the Dirac basis vector expansions of in-state
vectors $\phi^+$ as\footnote{We assume the absence of bound states for
  simplicity.}
\begin{gather}\label{24}
\phi^{+}=\sum_{j,j_{3},\eta}\int_{0}^{\infty}  \vert
E,j,j_{3},\eta^{+} \rangle\langle^{+}
E,j,j_{3},\eta\vert\phi^{+}\rangle\, dE = \int_0^\infty
|E^+\rangle\langle^+E|\phi^+\rangle\,dE
\end{gather}
and vectors $|E^-\rangle$ are taken as basis vectors for the Dirac
expansion of the out-vectors for the observables
$|\psi\rangle\langle\psi|$:
\begin{gather}\label{25}
\psi^{-}=\sum_{j,j_{3},\eta}\int_{0}^{\infty}  \vert
  E,j,j_{3},\eta^{-} \rangle\langle^{-}
  E,j,j_{3},\eta\vert\psi^{-}\rangle\,dE =
  \int_0^\infty |E^-\rangle\langle ^-E|\psi^-\rangle\,dE .
\end{gather}
The expansions~\eqref{24}, \eqref{25} are statement of the nuclear
spectral theorem for the pair of the
RHS's~\eqref{eq:16pk}~\cite{G,M,BG,CG}.

As was stated previously, $j$, $j_3$, $\eta$ denote the additional
quantum numbers. Here $j$, $j_3$ have been chosen to correspond to the angular momentum and
$\eta$ to particle species labels, e.g., channel quantum numbers
etc. The Dirac basis vector expansions (\ref{24}) and (\ref{25}) use
two dif\/ferent kinds of kets: $|E^\mp\rangle= \vert
Ejj_3\eta^\mp\rangle\in\Phi_\pm^\times$, as suggested by~\eqref{23},
the Lippmann--Schwinger out-plane waves $\vert E^-\rangle$ and
in-plane waves $\vert E^+\rangle$, respectively. These kets should
belong to the duals in certain Gel'fand triplets~\eqref{eq:16pk}.

The properties of the Ge'lfand triplets~\eqref{eq:16pk} are def\/ined by the mathematical properties of the set of the energy wave functions in the same way as the Schwartz space~$\Phi$ of~\eqref{19} is def\/ined by the set of energy wave functions  ${\cal S}_{{\mathbb R}^+}=\{\phi(E)\}$ in~\eqref{20}. The properties of the energy wave functions $\phi^{+}(E)=\langle^{+}E\vert\phi^{+}\rangle$ in~\eqref{24} and of $\psi^{-}(E)=\langle^{-}E\vert\psi^{-}\rangle$ in~\eqref{25} have been obtained from the phenomenology of resonance and decay phenomena~\cite{BQM}. From these properties the functions
\begin{gather}\label{26}
\phi^+(E)\equiv\langle^+E|\phi^+\rangle=\langle^+Ejj_3\eta\vert\phi^+\rangle=
\langle\phi^+\vert Ejj_3\eta^+\rangle^*
\end{gather}
were identified as the boundary value on the positive semiaxis ${\mathbb R}^+$ of an
\textbf{Hardy function in the lower complex energy half-plane}~\cite{PK1}. It is
convenient to def\/ine this analytic function on the lower half plane
which corresponds to the second sheet of the Riemann surface that
supports the values of the $S$-matrix $S_j(z)$.

Similarly, the energy wave function of the
observable $\vert\psi^-\rangle\langle\psi^-\vert$
\begin{gather}\label{27}
\psi^-(E)\equiv
\langle^-E|\psi^-\rangle=\langle^-Ejj_3\eta\vert\psi^-\rangle
\end{gather}
can be extended into an \textbf{Hardy function in the upper complex
  energy half-plane}~\cite{PK1} on the second sheet of the Riemann surface
corresponding to the $S$-matrix. Then $\langle\psi^{-}\vert E^{-}\rangle\langle^{+}E
\vert\phi^{+}\rangle$ in the $S$-matrix element~\eqref{29} below can
be continued into the lower complex energy semi-plane (2-nd sheet of
the $S$-matrix).

The situation can be summarized in the following diagram:
\begin{center}
\begin{tabular}{llll}
Two sets of Hardy functions & $\begin{array}{l}\text{for the
two}\\ \text{sets of vectors}\end{array}$ &
$\begin{array}{l}\text{from two sets}\\\text{of L-Sch.
kets}\end{array}$
& $\begin{array}{l}\text{leading to two}\\\text{Hardy spaces}\end{array}$\\[3ex]
$\{\phi^+(z)\!=\!\langle^+z|\phi^+\rangle\}$ & $\{\phi^+\}\!=\! {\rm states}$ &
$|E^+\rangle\!=\!|E,j,j_3,\eta^+\rangle$
 & \multicolumn{1}{c}{$\Phi_-$}\\[2ex]
$\{\psi^-(z)\!=\!\langle^-z|\psi^-\rangle\!=\!{\langle\psi^-|z^-\rangle}^{*}\}$
& $\{\psi^-\}\!=\! {\rm observables}$ & $|E^-\rangle\!=\!|E,j,j_3,\eta^-\rangle$ &
\multicolumn{1}{c}{$\Phi_+$}
\end{tabular}
\end{center}

From the analyticity property follows that $\langle\psi^{-}\vert
E^{-}\rangle \langle E^{+}\vert\psi^{+}\rangle S_{j}(E)$ can be
analytically continued into the lower complex semi-plane (second
sheet), except for singularities of $S_{j}(E)$.

{\sloppy This means that the contour of integration of the $S$-matrix element
\begin{gather}
(\psi^{-},\phi^{+})=
  (\psi^{\text{out}},S\phi^{\text{in}})
  =\sum_{j}\int_0^{\infty}\!\! dE \sum_{j_{3}}\sum_{\eta,\eta'}
  \langle\psi^{-}\vert E,j,j_{3},\eta'^{-}\rangle S_{j}^{\eta'\eta}(E)
  \langle^{+}E,j,j_{3},\eta\vert\phi^{+}\rangle.\!\!\!\label{29}
\end{gather}
can be deformed into the lower complex energy plane, second sheet of
the $S$-matrix ele\-ment~$S_{j}(E)$.

}

\begin{figure}[t]
  \centering
\includegraphics[width=110mm]{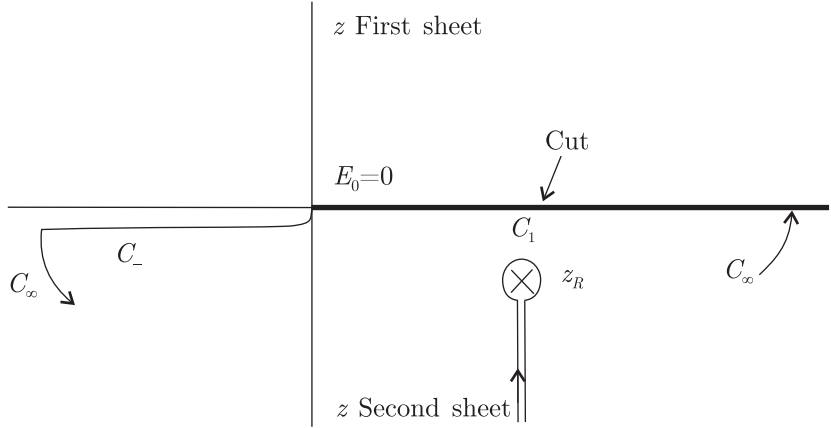}
  \caption{The second sheet of the the complex energy Riemann surface
    for the $S$-matrix with one resonance pole.}\label{Fig1}
\end{figure}

Consider the simplest case that there is one $S$-matrix pole at
$z_{R}=E_{R}-i\Gamma/2$ on the second sheet as shown in Fig.~\ref{Fig1}. The
program to determine the property of the spaces $\Phi_{-}$ (states)
and~$\Phi_{+}$ (observables) is the following:

Start with the integral~\eqref{29} along the cut. In simplif\/ied
notation the $S$-matrix element~\eqref{29} is given by
\begin{gather}\label{30}
(\psi^{-},\phi^{+})=\int_{0}^{\infty} dE\langle\psi^{-}\vert
E^{-}\rangle S_{j}(E)\langle^{+} E\vert\phi^{+}\rangle .
\end{gather}
The contour of integration is now deformed from the positive real
axis f\/irst sheet through the cut into the lower complex energy plane
in the \textit{second} sheet, where the resonance pole of the
$S$-matrix is located at~$E=z_{R}$. Then one obtains the integrals
along~$\mathcal{C}_{-}$, along the semicircle~$\mathcal{C}_\infty$
and the integral $\mathcal{C}_{1}$ around the pole at $z_{R}$. This
is depicted in Fig.~\ref{Fig1}. The integral along~$\mathcal{C}_{\infty}$
vanishes and the integral along~$\mathcal{C}_{-}$ gives some
non-resonant background, whereas we wish to determine the needed
properties of the energy wave functions~\eqref{26}, \eqref{27} from
the resonance and decay properties associated to the $S$-matrix pole
at~$z_{R}$.

The task is: conjecture the mathematical property of
$\langle^{-}E\vert \psi^{-}\rangle$, $\langle^{+}E\vert
\phi^{+}\rangle$ such that a scat\-te\-ring resonance and a decaying
state is the same physical entity derived from the $S$-matrix pole
at~$z_{R}$ on the second sheet.

\subsection{Conjecturing the Hardy space axiom}\label{section3.3}

We want to f\/ind the mathematical def\/inition of the spaces
$\{\phi^{+}\}$ and $\{\psi^{-}\}$. For this purpose we start with
the $S$-matrix pole def\/inition of a resonance at
$z_{R}=E_{R}-i\Gamma/2$ on the second sheet of the $j$-th partial
$S$-matrix element $S_{j}(E)$.

We obtain (using the Cauchy theorem) from the pole term of
$S_{j}^{\text{P.T.}}(E)=2ia_{j}^{\text{BW}}(E)$ around~$z_{R}${\samepage
\begin{subequations}\label{eq:17pk}
{\addtolength{\tabcolsep}{-4pt}
\begin{tabular}{@{}lcl@{}}
  a Breit--Wigner resonance&& and
  related to it a Gamow state vector $\vert z_{R},j,j_{3},\eta\rangle$:\tsep{4pt}\\
amplitude $a_{j}^{\text{BW}}(E)$& \\
\begin{minipage}[b][17pt]{0.35\linewidth}
  \begin{equation}\label{eq:17apk}
    a_{j}^{\text{BW}}(E)=\frac{R}{E-E_{R}-i\Gamma/2}\!
\end{equation}
\end{minipage}
&
    $\Leftrightarrow$
&
\begin{minipage}[b][17pt]{0.6\linewidth}
\begin{equation}\label{eq:17bpk}
    \vert z_{R},j,j_{3},\eta\rangle\sqrt{2\pi\Gamma}
    =\int_{-\infty}^{+\infty}\!\!dE \vert E,j,j_{3},\eta^{-}
    \rangle\frac{i\sqrt{\frac{\Gamma}{2\pi}}}{E-z_{R}}.\!
\end{equation}
\end{minipage}
\end{tabular}}
\end{subequations}
}

\medskip

\noindent
For the Gamow ket one can show that it is an eigenket of $H$ with a
discrete complex eigenvalue (as Gamow wanted):
\begin{subequations}
\begin{gather}
  \label{eq:18pk}
  H^{\times}\vert E_{R}-i\Gamma/2,j,j_{3},\eta^{-}\rangle
  =(E_{R}-i\Gamma/)\vert E_{R}-i\Gamma/2,j,j_{3},\eta^{-}\rangle.
\end{gather}
Further one can show that the Gamow ket has the property:
\begin{gather}
  \langle\text{e}^{iHt/\hbar}\psi_{\eta}^{-} \vert
  E_{R}-i\Gamma/2,j,j_{3},\eta^{-}\rangle =
\langle\psi_{\eta}^{-} \vert \text{e}^{-iH^{\times}t/\hbar}\vert
  E_{R}-i\Gamma/2,j,j_{3},\eta^{-}\rangle\nonumber\\
\qquad{}= \text{e}^{-iE_{R}t/\hbar}\text{e}^{-(\Gamma/2)t/\hbar}
\langle\psi_{\eta}^{-} \vert
  E_{R}-i\Gamma/2,j,j_{3},\eta^{-}\rangle\qquad\text{but for
  }t\geq0 \ (=t_{0}) \ \ \text{only}.  \label{eq:19pk}
\end{gather}
\end{subequations}
This proves that the Breit--Wigner resonance observed by the
Breit--Wigner cross section
\begin{gather}\label{26pk}
  \sigma_{j}^{R}(E)=\frac{4\pi}{p^{2}}(2j+1)
  \frac{(\Gamma/2)^{2}}{(E-E_{R})^{2}+(\Gamma/2)^{2}}
\end{gather}
(i.e., by the lineshape~\eqref{26pk} as a function of $E$), is the same
physical object as the exponentially decaying state~\eqref{eq:19pk},
observed by the exponential time evolution:
\begin{gather}
  \label{eq:20pk}
  \vert  \langle\text{e}^{iHt/\hbar}\psi_{\eta}^{-} \vert
  E_{R}-i\Gamma/2,j,j_{3},\eta^{-}\rangle\vert^{2}
=\text{e}^{-\Gamma t/\hbar}
\vert \langle\psi_{\eta}^{-} \vert
  E_{R}-i\Gamma/2,j,j_{3},\eta^{-}\rangle \vert^{2}.
\end{gather}
This means that the ket~\eqref{eq:17bpk} with Breit--Wigner resonance
distribution is a Gamow ket~\eqref{eq:18pk} with exponential
time evolution~\eqref{eq:20pk}:
\begin{center}
  Breit--Wigner resonance of width $\Gamma$  $\equiv$ a decaying state
  with lifetime $\tau=\hbar/\Gamma$.
\end{center}
Therewith we have a theory that unif\/ies the concept of a
Breit--Wigner resonance~\eqref{eq:17apk}, \eqref{26pk} and the concept of an
exponentially decaying Gamow states~\eqref{eq:17bpk}, \eqref{eq:20pk}.

In order to separate the pole term in~\eqref{eq:17apk} from the $S$-matrix
in~\eqref{30} and associate to it a~ket $\vert
z_{R}=(E_R-i\Gamma/2),j,j_3,\eta\rangle$ in~\eqref{eq:17bpk} with a
Breit--Wigner energy distribution extending over $-\infty<E<+\infty$
(and not over the spectrum $E_{0}=0\leq E<\infty$ of the Hilbert space
opera\-tor~$H$ in~\eqref{24} and~\eqref{25}), and in order to
derive~\eqref{eq:18pk}, \eqref{eq:19pk} and~\eqref{eq:20pk} as generalized
eigenvector equations, new mathematical conditions must be met by
the energy wave functions $\langle^{-}E\vert\psi^{-}\rangle$ and
$\langle^{+}E\vert\phi^{+}\rangle$ of~\eqref{26}, \eqref{27} in
addition to the usual analyticity properties in the upper and lower
complex semi-plane, and in addition to their Schwartz space
property\footnote{Inf\/initely
  dif\/ferentiable and rapidly decreasing.} on the positive real axis~$\mathbb{R}$.

From these mathematical conditions the energy wave functions were
identif\/ied~\cite{PK1} as Hardy functions of the complex energy
semi-planes and this hypothesis was discussed subsequently in several publications~\cite{PK2,T,PK3,PK4}.

Specif\/ically, the energy wave functions of a state must be smooth
Hardy functions on the lower complex plane $\mathcal{C}_{-}$. In our
context this lower complex plane is taken as the second sheet of the
Riemann surface of the $S$-matrix:
\begin{gather*}
\phi^+(E) = \langle^+E|\phi^+\rangle \in (\mathcal{H}^2_-\cap {\cal
S})\Big|_{\mathbb{ R_+}}\,\equiv\text{Hardy functions on the lower
semi-plane }\mathcal{C}_{-}.
\end{gather*}

The energy wave functions of an observable are smooth Hardy functions
analytic on the upper semi-plane $\mathcal{C}_{+}$ (so that their
complex conjugates are analytic on the lower semi-plane):
\begin{gather*}
\psi^-(E) = \langle^-E|\psi^-\rangle \in (\mathcal{H}^2_+\cap
\mathcal{S})\Big|_{\mathbb{R_+}}\,\equiv\text{Hardy functions on the upper
semi-plane }\mathcal{C}_{+}.
\end{gather*}
Therewith we have inferred a new axiom of a causal quantum theory
which provides a mathematical description that unif\/ies resonance and
decay phenomena:
$$
  \textbf{Hardy space axiom}
$$
The set of prepared (in-) states def\/ined by the preparation
apparatus (e.g.\ accelerator) is described by the triplet
\begin{gather}\label{43}
\{\phi^+\}\equiv\Phi_- \subset \mathcal{H} \subset \Phi_{-}^{\times}.
\end{gather}
The set of (out-) observables def\/ined by the registration
apparatus (e.g.\ detector) is described by the triplet
\begin{gather}\label{44}
\{\psi^-\}\equiv\Phi_+\subset \mathcal{H} \subset \Phi_{+}^{\times}.
\end{gather}
The spaces $\Phi_{\pm}$ are the Hardy spaces on the semi-planes
$\mathcal{C}_{\pm}$ (second sheet of the analytic $S$-matrix), and
$\Phi_{\pm}^{\times}$ are their respective
duals\footnote{$\Phi_{\pm}^{\times}$ are spaces of antilinear
  continuous functionals on $\Phi_{\pm}$.}.

The Lippmann--Schwinger kets $\vert E,j,j_{3},\eta^{\pm}\rangle$ the
Gamow kets $\vert R_{R}-i\Gamma/2,j,j_{3},\eta^{-}\rangle$
and\linebreak
$\vert z,j,j_{3},\eta^{\pm}\rangle$ are mathematically well def\/ined as
functionals on the space $\Phi_{\mp}$.

To distinguish in the mathematical description between states
$\{\phi^{+}\}\equiv\Phi_{-}$ and the observables
$\{\psi^{-}\}\equiv\Phi_{+}$ is quite natural, since
experimentalists distinguish between preparation apparatus (e.g.,
accelerator) for states and registration apparatus (detector) for
observables.

Having discovered the Hardy space axiom for the quantum theory of
resonances and decaying states one can derive mathematical
consequences of this axiom.

In the same way as the unitary group evolution followed by the
Stone--von~Neumann theorem from the Hilbert space axiom, the time
evolution for the Hardy space axiom follows from a~mathematical
theorem for Hardy spaces, the
Paley--Wiener theorem (1934)~\cite{PW}:

The solutions of the dynamical equations, the Schr\"odinger equation
for the state $\phi^+(t)$ and the Heisenberg equation for the
observable $\psi^-(t)$ are given by the semigroup
\begin{gather}
\label{p}
\begin{array}[c]{@{}ccc}
  \text{(Heisenberg equation):}  & \quad &
  \text{(Schr\"{o}dinger equation):}\\
  \text{of observables in space }\Phi_+ & \, & \text{for states in the
    space } \Phi_-\\[1ex]
  \psi^-(t)=e^{iHt/\hbar} \psi^- & \quad & \phi^+(t) =
  e^{-iHt/\hbar} \phi^+\\[1ex]
  t_0=0 \le t < \infty & \, & t_0=0 \le t < \infty.
\end{array}
\end{gather}

Therefore the probability for the time evolved observable
$\psi^-(t)$ in the state $\phi^+$ can now be predicted as the Born
probability:
\begin{gather*}
\mathcal{P}_{\phi^+}(\psi^-(t))= |\langle\psi^-(t)|\phi^+\rangle|^2
= |\langle
e^{iHt/\hbar}\psi^-|\phi^+\rangle|^2\nonumber\\
\phantom{\mathcal{P}_{\phi^+}(\psi^-(t))}{} =|\langle\psi^-|e^{-iH_+
t/\hbar} \phi^+\rangle|^2\qquad \text{for} \ \ t\ge t_0=0 \ \ \text{only}.
\end{gather*}
This is in agreement with experimental data for the detector counts
because
\begin{gather*}
  \mathcal{P}_{\phi^+}(\psi^-(t))\sim N(t)/N\text{ can be measured
    only for }t\geq t_{0}=0,
\end{gather*}
where $t_{0}$ is the time at which the state $\phi^{+}$ had been
prepared (causality).

\section{Concluding remarks}\label{section4}

In order to derive from the resonance pole of the $S$-matrix $S_j(E)$
at $z_R=E_R-i\Gamma/2$ the Breit--Wigner resonance amplitude
in~\eqref{eq:17apk} and to derive for it the Gamow ket
in~\eqref{eq:17bpk} with the property~\eqref{eq:18pk}, \eqref{eq:19pk}
such that the relation $\tau=\hbar/\Gamma$ is fulf\/illed, one has to
make various mathematical assumptions for the energy wave functions.
These mathematical assumptions go beyond the requirement that
$\phi^+(E)$ and $\psi^-(E)$ are the Schwartz functions of the
mathematical version of the Dirac formalisms and can be summarized by
the new Hardy space axiom~(\ref{43}),~(\ref{44}).

Starting with (\ref{43}) and (\ref{44}) as the new axiom for the
space of states $\{\phi^+\}=\Phi_-$ and the space of observables
$\{\psi^-\}=\Phi_+$ one just needs to use the mathematical theorem
of Paley and Wiener \cite{PW} to obtain a semigroup time evolution
(\ref{p}). For the special case of Gamow kets, one obtains from this
new axiom the exponential decay law (\ref{eq:20pk}) for the states
described by the Gamow ket. But one obtains \eqref{eq:19pk},
\eqref{eq:20pk} for $t>t_0=0$. This is an intuitively plausible
result if $t_0=0$ is the time at which the Gamow state
$\phi^G=\sqrt{2\pi\Gamma}\vert z_R,j,j_3,\eta\rangle$ is prepared.

An interesting question is the interpretation of the time $t_0=0$,
which could be any {\it finite} time in our life or the life of the
Universe, but certainly should not be a time before the big bang.
Since quantum theory deals with an ensemble of quanta, one would
also expect that the ``quantum mechanical beginning of time'' $t_0$
will be observed by an ensemble of times
$\{t_0^{(1)},t_0^{(2)},\dots\}$. How to interpret this ensemble of
times and how to identify these times $t_0^{(i)}$ in experiments is
an interesting question.

\appendix

\pdfbookmark[1]{Appendix: Hardy functions on a half plane}{appendix}
\section*{Appendix: Hardy functions on a half plane}

Let us consider the open upper half plane ${\mathbb
  C}^+:=\{z\in{\mathbb C}\;|\; \operatorname{Im}z> 0\}$, where $\mathbb
C$ is the f\/ield of complex numbers. A {\it Hardy function}
\cite{BG,CG,PW,Dur} $f(z)$ on the upper half plane is an analytic function
on~${\mathbb C}^+$ such that for any line $y=x+i\alpha$, $\alpha>0$
and constant (i.e., any parallel to the real axis in~${\mathbb C}^+$),
one has that
\begin{gather}\label{54}
  \int_{-\infty}^\infty |f(x+i\alpha)|^2\,dx<K<\infty.
\end{gather}
This $K$ is the same for all values of $\alpha>0$. As a consequence,
the function $f(x):=\lim\limits_{\alpha\mapsto 0}f(x+i\alpha)$ is well
def\/ined (almost elsewhere) on the real line~$\mathbb R$ and is square
integrable, i.e.,
\begin{gather*}
  \int_{-\infty}^\infty |f(x)|^2\,dx<K<\infty,
\end{gather*}
where this $K$ is the same we have in (\ref{54}). Thus, $f(x)$ is the
function given by the boundary values on the real axis of $f(z)$ and
therefore it is uniquely determined by $f(z)$. Conversely, a~theorem
due to Titchmarsh~\cite{Dur}, based on the Cauchy theorem, shows that we
can recover the values of~$f(z)$, $z\in{\mathbb C}^+$, once we know
the function~$f(x)$. Thus, a Hardy function on the upper half plane~$f(z)$ is uniquely determined by the function given its boundary
values on the real~$f(x)$ line and therefore, we can identify~$f(z)$
and~$f(x)$.

Let us call ${\cal H}_+^2$ the set of Hardy functions on the upper
half plane. Then, ${\cal H}_+^2$ has the following properties:
\begin{enumerate}\itemsep=0pt
\item[$i)$] ${\cal H}_+^2$ is a linear space. Since each $f(x)\in {\cal
    H}_+^2$ is square integrable, ${\cal H}_+^2$ is a subspace of~$L^2({\mathbb R})$.

\item[$ii)$] Assume that $f(x)\in L^2({\mathbb R})$. How can we
  recognize that $f(x)$ is a function in ${\cal H}_+^2$? For that, we
  have the Paley--Wiener theorem \cite{PW}, which states that $f(x)\in
  {\cal H}_+^2$ if and only if it is the Fourier transform of a square
  integrable function supported on (i.e., that vanishes outside of)
  the negative semiaxis ${\mathbb R}^-\equiv (-\infty,0]$.
\item[$iii)$] The boundary values on an interval of the real line of a
  Hardy function on the upper half plane uniquely determine such a
  function, as it happens for any complex analytic function on the
  open upper half plane. This is particularly true when this interval
  is the positive semiaxis ${\mathbb R}^+\equiv [0,\infty)$. In this
  case, we have a formula that recovers all the values of $f(z)\in
  {\cal H}_+^2$ for $z$ on the closed upper half plane (including the
  values on the negative semiaxis) from the values of $f(x)$ for
  $x>0$.  Furthermore, it gives a criteria that states when a square
  integrable function on the positive semiaxis can be continued into a
  Hardy function in ${\cal H}_+^2$. Both results rely on the
  properties of the Mellin transform \cite{VW}.
\end{enumerate}

Hardy functions on the lower half plane ${\mathbb C}^-\equiv
\{z\in{\mathbb C}\;| \; \operatorname{Im}z<0\}$ are def\/ined analogously.
The space of Hardy functions on the lower half plane are denoted by
${\cal H}_-^2$ and have similar properties than those above stated
for functions in ${\cal H}_+^2$, with minor dif\/ferences. In
particular, in $ii)$ we have to replace Fourier transforms of
function supported on the negative semiaxis by Fourier transforms
supported on the positive semiaxis~\cite{PW}.

There are some additional properties that concerns functions in both
Hardy spaces ${\cal H}_\pm^2$:
\begin{enumerate}\itemsep=0pt
\item[1.] Both spaces ${\cal H}_\pm^2$ are closed subspaces of
  $L^2({\mathbb R})$ and therefore Hilbert spaces with the scalar
  product of functions in $L^2({\mathbb R})$. Moreover, each function
  in ${\cal H}_\pm^2$ is orthogonal to each function in ${\cal
    H}_\mp^2$. In addition,
\begin{gather}\label{56}
  L^2({\mathbb R})={\cal H}_+^2\oplus {\cal H}_-^2,
\end{gather}
where the symbol $\oplus$ means orthogonal direct sum. This is indeed
a consequence of the Paley--Wiener theorem~\cite{PW}.
\item[2.] The complex conjugate $f^*(x)$ of a function $f(x)\in {\cal
    H}_\pm^2$ is in ${\cal H}_\mp^2$, $f(x)\in {\cal H}_\mp^2$.
  Moreover, $[f(z^*)]^*=f(z)$, for any $f(z)\in {\cal H}_\mp^2$, where
  the star denotes complex conjugation.
\item[3.] Let us call ${\cal H}_\pm^2\big|_{{\mathbb R}^+}$ and
  ${\cal H}_\pm^2\cap{\cal S}\big|_{{\mathbb R}^+}$ the spaces of the
  restrictions to the positive semiaxis~${\mathbb R}^+$ of the
  functions in ${\cal H}_\pm^2$ or in the intersection ${\cal
    H}_\pm^2\cap{\cal S}$ of the Hardy spaces with the Schwartz space,
  respectively. These two spaces are dense in $L^2({\mathbb R}^+)$
  \cite{BG}.
\end{enumerate}

\subsection*{Acknowledgments}

We wish to acknowledge partial f\/inancial support by the Spanish
Ministry of Science and Innovation through Project MTM2009-10751,
the Junta de Castilla y Le\'on, through Project GR224 and the US NSF
Award no OISE-0421936.

\pdfbookmark[1]{References}{ref}
\LastPageEnding

\end{document}